\documentstyle[amssymb,floats,aps,prb,epsf]{revtex}
\epsfclipon
\begin{document}
\twocolumn[\hsize\textwidth\columnwidth\hsize\csname
@twocolumnfalse\endcsname

\draft
\draft
\title{Asymmetry of the electron spectrum in hole-doped and
electron-doped cuprates }
\author{Huaiming Guo and Shiping Feng}
\address{Department of Physics, Beijing Normal University,
Beijing 100875, China}

\maketitle
\begin{abstract}
Within the $t$-$t'$-$J$ model, the asymmetry of the electron
spectrum and quasiparticle dispersion in hole-doped and
electron-doped cuprates is discussed. It is shown that the
quasiparticle dispersions of both hole-doped and electron-doped
cuprates exhibit the flat band around the $(\pi,0)$ point below
the Fermi energy. The lowest energy states are located at the
$(\pi/2,\pi/2)$ point for the hole doping, while they appear at
the $(\pi,0)$ point in the electron-doped case due to the
electron-hole asymmetry. Our results also show that the unusual
behavior of the electron spectrum and quasiparticle dispersion is
intriguingly related to the strong coupling between the electron
quasiparticles and collective magnetic excitations.

\end{abstract}

]
\bigskip

\narrowtext The parent compounds of cuprate superconductors are
believed to belong to a class of materials known as Mott
insulators with an antiferromagnetic (AF) long-range order
(AFLRO), then superconductivity emerges when charge carriers,
holes or electrons, are doped into these Mott insulators
\cite{kastner,bednorz,tokura}. Although both hole-doped and
electron-doped cuprates have the layered structure of the square
lattice of the CuO$_{2}$ plane separated by insulating layers
\cite{kastner,bednorz,tokura}, the significantly difference of the
electronic states between hole-doped and electron-doped cuprates
is observed \cite{shen,homes}, which reflects the electron-hole
asymmetry. For the hole-doped cuprates, AFLRO is reduced
dramatically with doping \cite{kastner,yamada}, and vanished
around the doping $\delta\sim 0.05$. But a series of inelastic
neutron scattering measurements show that the incommensurate short
AF correlation persists in the underdoped, optimally doped, and
overdoped regimes \cite{kastner,yamada}, then in low temperatures,
the systems become superconducting (SC) over a wide range of the
hole doping concentration $\delta$, around the optimal doping
$\delta\sim 0.15$ \cite{bednorz,tallon}. However, AFLRO survives
until superconductivity appears over a narrow range of $\delta$,
around the optimal doping $\delta\sim 0.15$ in the electron-doped
cuprates \cite{tokura,tokura1,peng}. In particular, the maximum
achievable SC transition temperature in the electron-doped
cuprates is much lower than that in the hole-doped case, and the
commensurate spin response in the SC-state is observed
\cite{yamada1}. These experimental observations show that the
unconventional physical properties of both hole-doped and
electron-doped cuprates mainly depend on the extent of the doping
concentration. Since many of the unconventional physical
properties, including the relatively high SC transition
temperature, have often been attributed to particular
characteristics of low energy excitations determined by the
electronic structure \cite{kastner,shen}, then a central issue to
clarify the nature of the unconventional physical properties is
how the electronic structure evolves with the doping
concentration,

From the angle-resolved photoemission spectroscopy (ARPES)
measurements \cite{shen,kim}, it has been shown that the electron
spectral function $A({\bf k},\omega)$ in doped cuprates is
strongly momentum and doping dependent. For the hole doping, the
charge carriers doped into the parent Mott insulators first enter
into the ${\bf k}=[\pi/2,\pi/2]$ (in units of inverse lattice
constant) point  in the Brillouin zone
\cite{shen,kim,dessau,wells}, while the charge carriers are
accommodated at the ${\bf k}=[\pi,0]$ point in the electron-doped
case \cite{shen,kim,armitage,kim1}. Moreover, $A({\bf k},\omega)$
has a flat band form as a function of energy $\omega$ for ${\bf
k}$ in the vicinity of the $[\pi,0]$ point, which leads to the
unusual quasiparticle dispersion around the $[\pi,0]$ point with
anomalously small changes of electron energy as a function of
momentum \cite{shen,kim,dessau,wells,armitage,kim1}. In
particular, this flat band is just below the Fermi energy for the
hole-doped cuprates, while it is located well below the Fermi
energy for the electron-doped case. The flat band reflects the
underlying electronic structure near a band saddle point, and is
manifestation of a strong coupling between the electron
quasiparticles and collective excitations \cite{shen3}. Since the
normal-state pseudogap starts growing first in the single-particle
excitations around the $[\pi,0]$ point, and then exists in a wide
range of the doping concentration \cite{shen}, therefore the broad
feature in the electron spectrum around the $[\pi,0]$ point has a
particular importance in the mechanism of the normal-state
pseudogap formation, and is responsible for the unconventional
normal-state properties \cite{shen}. Recently, a new low photon
energy regime of ARPES has been accessed with lasers and used to
study cuprate superconductors, where the clearest evidence for the
existence of the electron quasiparticles in the normal-state has
been observed \cite{koralek}. Therefore these ARPES experiments
have produced some interesting data that introduce important
constraints on the model and theory of cuprate superconductors
\cite{shen}.

The doping evolution of the normal-state electron spectrum in the
hole-doped cuprates has been extensively studied within some
strongly correlated models \cite{dagotto}. The most striking
aspect is that the unusual quasiparticle dispersion can not be
explained by either of the band theory scenarios. The numerical
calculation of the electron spectrum based on the large U Hubbard
\cite{bulut} model for the hole-doped case shows the quasiparticle
dispersion similar to those observed in experiments. In this
calculation, the flat band around the $[\pi,0]$ point arises from
the large Coulomb interaction U. The unusual quasiparticle
dispersion in the hole-doped case has been studied numerically
within the $t$-$J$ type model \cite{dagotto1}, and the result of
the quasiparticle dispersion along the $[0,0]$ to $[\pi,0]$
direction is quantitative agreement with the experiments. This
study also shows unambiguously that the energy scale of the
quasiparticle band is controlled by the magnetic interaction $J$.
In particular, the electron-hole asymmetry in hole-doped and
electron-doped cuprates has been discussed based on numerically
exact diagonalization methods \cite{gooding}, it is shown that the
electron-hole asymmetry comes from the coupling of the charge
carriers with the spin background. Moreover, many authors
\cite{anderson,laughlin} suggest that the unusual electron
spectrum in doped cuprates is a natural consequence of the
charge-spin separation (CSS). Although the exact origin of the
striking behavior of the electron spectrum still is controversial,
a strongly correlated many-body like approach may be appropriate
to describe the electronic structure of doped cuprates. Recently,
we \cite{feng1} have developed a CSS fermion-spin theory for
description of the unconventional physical properties in doped
cuprates, where the electron operator is decoupled as a gauge
invariant dressed charge carrier and spin. Within this framework,
we have shown that the charge transport is mainly governed by the
scattering from the dressed charge carriers due to the spin
fluctuation, and the scattering from the spins due to the dressed
charge carrier fluctuation dominates the spin response
\cite{feng1}. In particular, the charge-spin recombination of the
dressed charge carrier and spin automatically gives the electron
quasiparticle character \cite{feng2}. In this paper, we study the
asymmetry of the electron spectrum in hole-doped and
electron-doped cuprates along with this line. Our results show
that the quasiparticle dispersions of both hole-doped and
electron-doped cuprates exhibit the flat band around the $[\pi,0]$
point below the Fermi energy. The lowest energy states are located
at the $[\pi/2,\pi/2]$ point for the hole doping, while they
appear at the $[\pi,0]$ point in the electron-doped case due to
the electron-hole asymmetry. Our results also show that the
striking behavior of the electron spectrum in doped cuprates is
intriguingly related to the strong coupling between the electron
quasiparticles and collective magnetic excitations.

In both hole-doped and electron-doped cuprates, the characteristic
feature is the presence of the two-dimensional CuO$_{2}$ plane
\cite{bednorz,tokura} as mentioned above, and it seems evident
that the unusual behaviors are dominated by this CuO$_{2}$ plane.
Although the $t$-$J$ model captures the essential physics of the
doped CuO$_{2}$ plane \cite{anderson1}, the electron-hole
asymmetry may be properly accounted by generalizing the $t$-$J$
model to include the second-nearest neighbors hopping terms $t'$
\cite{hybertson}. In this case, we start from the $t$-$t'$-$J$
model on a square lattice,
\begin{eqnarray}
H&=&-t\sum_{i\hat{\eta}\sigma}C^{\dagger}_{i\sigma}
C_{i+\hat{\eta}\sigma} +t'\sum_{i\hat{\tau}\sigma}
C^{\dagger}_{i\sigma}C_{i+\hat{\tau}\sigma}\nonumber\\
&+&\mu_{0}\sum_{i\sigma}
C^{\dagger}_{i\sigma}C_{i\sigma}+J\sum_{i\hat{\eta}}{\bf S}_{i}
\cdot {\bf S}_{i+\hat{\eta}},
\end{eqnarray}
with $\hat{\eta}=\pm\hat{x},\pm\hat{y}$, $\hat{\tau}=\pm\hat{x}
\pm\hat{y}$, $C^{\dagger}_{i\sigma}$ ($C_{i\sigma}$) is the
electron creation (annihilation) operator, ${\bf S}_{i}=
C^{\dagger}_{i}{\vec\sigma}C_{i}/2$ is spin operator with
${\vec\sigma}=(\sigma_{x},\sigma_{y},\sigma_{z})$ as Pauli
matrices, and $\mu_{0}$ is the chemical potential. For the
electron-doped case, we can perform a particle-hole transformation
$C_{i\sigma}\rightarrow C^{\dagger}_{i-\sigma}$, so that the
difference between hole-doped and electron-doped cases is
expressed as the sign difference of the hopping parameters, i.e.,
$t>0$ and $t'>0$ for the hole doping and $t<0$ and $t'<0$ for the
electron doping \cite{feng3}. In this case, the $t$-$t'$-$J$ model
(1) in both hole-doped and electron-doped cases is always subject
to an important on-site local constraint to avoid the double
occupancy, i.e., $\sum_{\sigma} C^{\dagger}_{i\sigma}C_{i\sigma}
\leq 1$. Therefore the strong electron correlation in the
$t$-$t'$-$J$ model (1) manifests itself by this single occupancy
local constraint \cite{anderson1}. It has been shown that this
local constraint can be treated properly in analytical
calculations within the CSS fermion-spin theory \cite{feng1},
$C_{i\uparrow}= h^{\dagger}_{i\uparrow}S^{-}_{i}$ and
$C_{i\downarrow}= h^{\dagger}_{i\downarrow} S^{+}_{i}$, where the
spinful fermion operator $h_{i\sigma}= e^{-i\Phi_{i\sigma}}h_{i}$
describes the charge degree of freedom together with some effects
of the spin configuration rearrangements due to the presence of
the doped charge carrier itself (dressed charge carrier), while
the spin operator $S_{i}$ describes the spin degree of freedom,
then the local constraint for the single occupancy, $\sum_{\sigma}
C^{\dagger}_{i\sigma}C_{i\sigma}=S^{+}_{i} h_{i\uparrow}
h^{\dagger}_{i\uparrow}S^{-}_{i}+ S^{-}_{i}h_{i\downarrow}
h^{\dagger}_{i\downarrow}S^{+}_{i}=h_{i} h^{\dagger}_{i}(S^{+}_{i}
S^{-}_{i}+S^{-}_{i}S^{+}_{i})=1- h^{\dagger}_{i}h_{i}\leq 1$, is
satisfied in analytical calculations. These dressed charge carrier
and spin are gauge invariant \cite{feng1}, and in this sense, they
are real and can be interpreted as the physical excitations
\cite{laughlin}. Although in common sense $h_{i\sigma}$ is not a
real spinful fermion, it behaves like a spinful fermion. In this
CSS fermion-spin representation, the low-energy behavior of the
$t$-$t'$-$J$ model (1) can be expressed as \cite{feng1},
\begin{eqnarray}
H&=&-t\sum_{i\hat{\eta}}(h_{i\uparrow}S^{+}_{i}
h^{\dagger}_{i+\hat{\eta}\uparrow}S^{-}_{i+\hat{\eta}}+
h_{i\downarrow}S^{-}_{i}h^{\dagger}_{i+\hat{\eta}\downarrow}
S^{+}_{i+\hat{\eta}})\nonumber\\
&+&t'\sum_{i\hat{\tau}}(h_{i\uparrow}S^{+}_{i}
h^{\dagger}_{i+\hat{\tau}\uparrow}S^{-}_{i+\hat{\tau}}+
h_{i\downarrow}S^{-}_{i}h^{\dagger}_{i+\hat{\tau}\downarrow}
S^{+}_{i+\hat{\tau}}) \nonumber \\
&-&\mu_{0}\sum_{i\sigma}h^{\dagger}_{i\sigma}h_{i\sigma}+J_{{\rm
eff}}\sum_{i\hat{\eta}}{\bf S}_{i}\cdot {\bf S}_{i+\hat{\eta}},
\end{eqnarray}
with $J_{{\rm eff}}=(1-\delta)^{2}J$, and $\delta=\langle
h^{\dagger}_{i\sigma}h_{i\sigma}\rangle=\langle h^{\dagger}_{i}
h_{i}\rangle$ is the doping concentration. As a consequence, the
magnetic energy ($J$) term in the $t$-$t'$-$J$ model is only to
form an adequate spin configuration \cite{anderson2}, while the
kinetic energy part has been expressed as the interaction between
the dressed charge carriers and spins, which reflects that even
the kinetic energy part in the $t$-$t'$-$J$ Hamiltonian has the
strong Coulombic contribution due to the restriction of no doubly
occupancy of a given site, and therefore dominates the essential
physics in doped cuprates.

For the discussions of the electron spectrum, we need to calculate
the electron Green's function $G(i-j,t-t')=\langle\langle
C_{i\sigma}(t);C^{\dagger}_{j\sigma}(t')\rangle\rangle$, which is
a convolution \cite{anderson2} of the spin Green's function
$D(i-j,t-t')=\langle\langle S^{+}_{i}(t);S^{-}_{j}(t')\rangle
\rangle$ and dressed charge carrier Green's function $g(i-j,t-t')
=\langle\langle h_{i\sigma}(t);h^{\dagger}_{j\sigma}(t')\rangle
\rangle$, and can be formally expressed in terms of the spectral
representation as,
\begin{mathletters}
\begin{eqnarray}
G_{{\rm p-type}}({\bf k},\omega)&=&{1\over N}\sum_{{\bf q}}
\int^{\infty}_{-\infty} {d\omega' \over 2\pi}
\int^{\infty}_{-\infty}{d\omega'' \over 2\pi} A_{h}({\bf q},
\omega')\nonumber\\
&\times&A_{s}({\bf q+k},\omega''){n_{F}(\omega')+
n_{B}(\omega'')\over \omega+\omega'-\omega''}, \\
G_{{\rm n-type}}({\bf k},\omega)&=&{1\over N}\sum_{{\bf q}}
\int^{\infty}_{-\infty} {d\omega' \over 2\pi}
\int^{\infty}_{-\infty}{d\omega'' \over 2\pi} A_{h}({\bf q},
\omega')\nonumber\\
&\times&A_{s}({\bf q+k},\omega''){1-n_{F}(\omega')+
n_{B}(\omega'')\over \omega-\omega'-\omega''},
\end{eqnarray}
\end{mathletters}
for hole-doped and electron-doped cases, respectively, where the
dressed charge carrier spectral function $A_{h}({\bf q},\omega)=
-2{\rm Im}g({\bf q},\omega)$, the spin spectral function
$A_{s}({\bf k},\omega)=-2{\rm Im}D({\bf k},\omega)$, and
$n_{B}(\omega)$ and $n_{F}(\omega)$ are the boson and fermion
distribution functions, respectively. This convolution reflects
the charge-spin recombination \cite{anderson2}. In the calculation
of the electron Green's function (3b) for the electron-case, the
particle-hole transformation $C_{i\sigma}\rightarrow
C^{\dagger}_{i-\sigma}$ as mentioned above has been considered.
Since the quantum spin operators obey the Pauli spin algebra,
i.e., the spin one-half raising and lowering operators $S^{+}_{i}$
and $S^{-}_{i}$ behave as fermions on the same site and as bosons
on different sites, then this problem can be discussed in terms of
the equation of motion method \cite{feng1,feng2}. It has been
shown that in the mean-field (MF) level, the spin system is an
anisotropic away from the half-filling \cite{feng2}. In this case,
we need to define another spin Green's function $D_{z}(i-j,t-t')=
\langle\langle S^{z}_{i}(t);S^{z}_{j}(t')\rangle\rangle$, and then
both spin Green's functions $D({\bf p},\omega)$ and $D_{z}({\bf
p},\omega)$ describe the spin propagations. In the doped regime
without AFLRO, i.e., $\langle S^{z}_{i}\rangle =0$, a MF theory of
the $t$-$J$ model has been developed \cite{feng2}. Following their
discussions, the MF dressed charge carrier and spin Green's
functions of the $t$-$t'$-$J$ model have been obtained as
\cite{feng1},
\begin{mathletters}
\begin{eqnarray}
g^{(0)}({\bf k},\omega)&=&{1\over \omega-\xi_{{\bf k}}}, \\
D^{(0)}({\bf p},\omega)&=&{B_{{\bf p}}\over 2\omega_{{\bf p}}}
\left ( {1\over \omega-\omega_{{\bf p}}}-{1\over\omega+
\omega_{{\bf p}}} \right ), \\
D^{(0)}_{z}({\bf p},\omega)&=&{B_{z{\bf p}}\over 2\omega_{z{\bf
p}}}\left ( {1\over \omega-\omega_{z{\bf p}}}-{1\over \omega +
\omega_{z{\bf p}}} \right ),
\end{eqnarray}
\end{mathletters}
where $B_{{\bf p}}=2\lambda_{1}(A_{1}\gamma_{{\bf p}}- A_{2})-
\lambda_{2}(2\chi^{z}_{2}\gamma_{{\bf p }}'-\chi_{2})$, $B_{z{\bf
p}}=\epsilon\chi_{1}\lambda_{1}(\gamma_{{\bf p}}-1)-\chi_{2}
\lambda_{2}(\gamma_{{\bf p}}'-1)$, $\lambda_{1}= 2ZJ_{eff}$,
$\lambda_{2}=4Z\phi_{2} t'$,  $A_{1}= \epsilon
\chi^{z}_{1}+\chi_{1}/2$, $A_{2} =\chi^{z}_{1}+\epsilon
\chi_{1}/2$, $\epsilon=1+2t\phi_{1} /J_{{\rm eff}}$, the spin
correlation functions $\chi_{1}=\langle S_{i}^{+}
S_{i+\hat{\eta}}^{-}\rangle$, $\chi_{2}=\langle S_{i}^{+}
S_{i+\hat{\tau}}^{-}\rangle$, $\chi^{z}_{1}=\langle S_{i}^{z}
S_{i+\hat{\eta}}^{z}\rangle$, and $\chi^{z}_{2}=\langle S_{i}^{z}
S_{i+\hat{\tau}}^{z}\rangle$, the dressed charge carrier's
particle-hole parameters $\phi_{1}=\langle h^{\dagger}_{i\sigma}
h_{i+\hat{\eta}\sigma}\rangle$ and $\phi_{2}=\langle
h^{\dagger}_{i\sigma}h_{i+\hat{\tau}\sigma}\rangle$, $\gamma_{{\bf
p}}=(1/Z)\sum_{\hat{\eta}}e^{i{\bf p}\cdot \hat{\eta}}$,
$\gamma'_{{\bf p}}=(1/Z)\sum_{\hat{\tau}} e^{i{\bf p}\cdot
\hat{\tau}}$, $Z$ is the number of the nearest neighbor or
second-nearest neighbor sites, while the MF dressed charge carrier
and spin excitation spectra are given by,
\begin{mathletters}
\begin{eqnarray}
\xi_{{\bf k}}&=&\varepsilon_{{\bf k}}-\mu_{0},\\
\omega^{2}_{{\bf p}}&=&\lambda_{1}^{2}[(A_{4}-\alpha\epsilon
\chi^{z}_{1}\gamma_{{\bf p}}-{1\over 2Z}\alpha\epsilon\chi_{1})
(1-\epsilon\gamma_{{\bf p}})\nonumber\\
&+&{1\over 2}\epsilon(A_{3}-{1\over 2} \alpha\chi^{z}_{1}-\alpha
\chi_{1}\gamma_{{\bf p}})(\epsilon-
\gamma_{{\bf p}})] \nonumber \\
&+&\lambda_{2}^{2}[\alpha(\chi^{z}_{2}\gamma_{{\bf p}}'-{3\over
2Z}\chi_{2})\gamma_{{\bf p}}'+{1\over 2}(A_{5}-{1\over 2}\alpha
\chi^{z}_{2})]\nonumber\\
&+&\lambda_{1}\lambda_{2}[\alpha\chi^{z}_{1}(1-\epsilon
\gamma_{{\bf p}})\gamma_{{\bf p}}' +{1\over
2}\alpha(\chi_{1}\gamma_{{\bf p}}'-C_{3})(\epsilon- \gamma_{{\bf
p}})\nonumber\\
&+&\alpha \gamma_{{\bf p}}'(C^{z}_{3}-\epsilon \chi^{z}_{2}
\gamma_{{\bf p}})-{1\over 2}\alpha\epsilon(C_{3}-
\chi_{2} \gamma_{{\bf p}})],\\
\omega^{2}_{z{\bf p}}&=&\epsilon\lambda^{2}_{1}(\epsilon A_{3}-
{1\over Z}\alpha\chi_{1}-\alpha\chi_{1}\gamma_{{\bf p}})
(1-\gamma_{{\bf p}}) \nonumber\\
&+&\lambda^{2}_{2}A_{5}(1-\gamma_{{\bf p}}')+\lambda_{1}
\lambda_{2}[\alpha\epsilon C_{3}(\gamma_{{\bf p}}+\gamma_{{\bf
p}}'-2)\nonumber \\
&+&\alpha\chi_{2}\gamma_{{\bf p}}(1-\gamma_{{\bf p}}')],
\end{eqnarray}
\end{mathletters}
respectively, where $\varepsilon_{{\bf k}}= Zt\chi_{1}
\gamma_{{\bf k}}-Zt'\chi_{2}\gamma'_{{\bf k}}$, $A_{3}=\alpha
C_{1}+(1-\alpha)/(2Z)$, $A_{4}=\alpha C^{z}_{1}+(1-\alpha)/(4Z)$,
$A_{5}=\alpha C_{2}+(1-\alpha)/(2Z)$, and the spin correlation
functions $C_{1}=(1/Z^{2}) \sum_{\hat{\eta},\hat{\eta'}}\langle
S_{i+\hat{\eta}}^{+} S_{i+\hat{\eta'}}^{-}\rangle$,
$C^{z}_{1}=(1/Z^{2}) \sum_{\hat{\eta},\hat{\eta'}}\langle
S_{i+\hat{\eta}}^{z} S_{i+\hat{\eta'}}^{z}\rangle$,
$C_{2}=(1/Z^{2}) \sum_{\hat{\tau},\hat{\tau'}}\langle
S_{i+\hat{\tau}}^{+} S_{i+\hat{\tau'}}^{-}\rangle$,
$C_{3}=(1/Z)\sum_{\hat{\tau}} \langle S_{i+\hat{\eta}}^{+}
S_{i+\hat{\tau}}^{-}\rangle$, and $C^{z}_{3}=(1/Z)
\sum_{\hat{\tau}}\langle S_{i+\hat{\eta}}^{z}
S_{i+\hat{\tau}}^{z}\rangle$. In order to satisfy the sum rule of
the correlation function $\langle S^{+}_{i}S^{-}_{i}\rangle=1/2$
in the case without AFLRO, the important decoupling parameter
$\alpha$ has been introduced in the calculation \cite{feng2},
which can be regarded as the vertex correction.

In the MF approximation, the electron spectrum of the hole-doped
cuprates has been discussed based on the $t$-$J$ model
\cite{feng2}, where the majority feature is that the MF intensity
peaks in the electron spectrum at the high symmetry points are
qualitatively consistent with the numerical simulations. However,
for the qualitative comparison with the experimental results of
both hole-doped and electron-doped cuprates
\cite{shen,kim,dessau,wells,armitage,kim1}, the electron spectrum
and overall quasiparticle dispersion should be studied beyond the
MF approximation, since they are associated with the fluctuation
of the dressed charge carriers and spin (then electrons). In the
following discussions, we limit the spin part to the first-order
(the MF level) since the charge transport can be well described at
this level \cite{feng1,feng3,feng4}. On the other hand, it has
been shown that there is a connection between the charge transport
and the quasiparticle dispersion around the $[\pi,0]$ point
\cite{shen,newns}. Within the CSS fermion-spin theory, we
\cite{feng1,feng3,feng4} have discussed the charge transport of
both hole-doped and electron-doped cuprates, and found that there
is no direct contribution to the charge transport from the spins,
although the strong correlation between the dressed charge
carriers and spins has been considered through the spin's order
parameters entering in the dressed charge carrier part. Therefore
we treat the dressed charge carrier part beyond the MF
approximation by considering the fluctuation. In this case, we
obtain the self-consistent equation in terms of the equation of
motion method that satisfied by the full dressed charge carrier
Green's function as \cite{mahan,feng5},
\begin{eqnarray}
g({\bf k},i\omega_{n})&=&g^{(0)}({\bf k},i\omega_{n})\nonumber \\
&+&g^{(0)}({\bf k},i\omega_{n})\Sigma^{(h)}({\bf k},i\omega_{n})
g({\bf k}, i\omega_{n}) ,
\end{eqnarray}
with the dressed charge carrier self-energy is evaluated from the
spin pair bubble as,
\begin{eqnarray}
\Sigma^{(h)}({\bf k},i\omega_{n})&=& {1\over N^{2}}\sum_{{\bf
p,p'}}\Lambda({\bf k,p,p'}){1\over \beta}\sum_{ip_{m}}g({\bf
p+k},ip_{m}+i\omega_{n})\nonumber \\
&\times& {1\over\beta}\sum_{ip'_{m}}D^{(0)}({\bf p'},ip'_{m})
\nonumber\\&\times&D^{(0)}({\bf p'+p},ip'_{m}+ip_{m}),
\end{eqnarray}
where $\Lambda({\bf k,p,p'})=[Zt\gamma_{{\bf p+p'+k}}- Zt'
\gamma_{{\bf p+p'+k}}']^{2}$. This self-energy function
$\Sigma^{(h)}({\bf k},\omega)$ renormalizes the MF dressed charge
carrier spectrum, and therefore it describes the quasiparticle
coherence. In particular, $\Sigma^{(h)}({\bf k},\omega)$ is not
even function. For the convenience of the discussions,
$\Sigma^{(h)}({\bf k},\omega)$ can be broken up into its symmetric
and antisymmetric parts as, $\Sigma^{(h)}({\bf k},\omega)=
\Sigma^{(h)}_{e}({\bf k},\omega)+\omega\Sigma^{(h)}_{o}({\bf k},
\omega)$, therefore both $\Sigma^{(h)}_{e}({\bf k},\omega)$ and
$\Sigma^{(h)}_{o}({\bf k},\omega)$ are even functions of $\omega$.
Now we define the quasiparticle coherent weight as
$Z^{-1}_{F}({\bf k},\omega)=1-\Sigma^{(h)}_{o}({\bf k},\omega)$,
then the full dressed charge carrier Green's function in Eq. (6)
can be written as,
\begin{eqnarray}
g({\bf k},\omega)={Z_{F}({\bf k},\omega)\over\omega-Z_{F}({\bf k},
\omega)[\xi_{{\bf k}}+\Sigma^{(h)}_{e}({\bf k},\omega)]} .
\end{eqnarray}
Since we only discuss the low-energy behavior of doped cuprates,
then the quasiparticle coherent weight can be discussed in the
static limit, i.e., $Z^{-1}_{F}({\bf k})=1-\Sigma^{(h)}_{o}({\bf
k},\omega)\mid_{\omega=0}$ and $\Sigma^{(h)}_{e}({\bf k})=
\Sigma^{(h)}_{e}({\bf k},\omega) \mid_{\omega=0}$. Although
$Z_{F}({\bf k})$ and $\Sigma^{(h)}_{e}({\bf k})$ still are a
function of ${\bf k}$, the wave vector dependence is unimportant.
It has been shown from ARPES experiments
\cite{shen,kim,dessau,wells} that in the normal-state, the lowest
energy states are located at the $[\pi/2,\pi/2]$ point for the
hole-doped cuprates, and the $[\pi,0]$ point in the electron-doped
case \cite{shen,kim,armitage,kim1}, which indicates that the
majority contribution for the electron spectrum comes from the
$[\pi/2,\pi/2]$ point for the hole doping, and the $[\pi,0]$ point
for the electron doping. In this case, the wave vector ${\bf k} $
in $Z_{F}({\bf k})$ and $\Sigma^{(h)}_{e}({\bf k})$ can be chosen
as $Z^{-1}_{F}=1-\Sigma^{(h)}_{o}({\bf k})\mid_{{\bf k}={\bf
k}_{h}=[\pi/2,\pi/2]}$ and $\Sigma^{(h)}_{e}=\Sigma^{(h)}_{e}
({\bf k})\mid_{{\bf k}={\bf k}_{h}=[\pi/2,\pi/2]}$ for the hole
doping, and $Z^{-1}_{F}=1-\Sigma^{(h)}_{o}({\bf k})\mid_{{\bf k}
={\bf k}_{e}=[\pi,0]}$ and $\Sigma^{(h)}_{e}=\Sigma^{(h)}_{e}
({\bf k})\mid_{{\bf k}={\bf k}_{e}= [\pi,0]}$ for the electron
doping, then the dressed charge carrier Green's function in Eq.
(8) can be expressed explicitly as,
\begin{eqnarray}
g({\bf k},\omega)&=&{Z_{F}\over \omega-\bar{\xi_{{\bf k}}}},
\end{eqnarray}
where the renormalized dressed charge carrier quasiparticle
spectrum $\bar{\xi}_{{\bf k}}=\bar{\varepsilon}_{{\bf k}}-\mu$,
with $\bar{\varepsilon}_{{\bf k}}=Z_{F}\varepsilon_{{\bf k}}$ and
renormalized chemical potential $\mu=Z_{F}(\mu_{0}-
\Sigma^{(h)}_{e})$. As we will see later, this $Z_{F}$ reduces the
dressed holon (then electron quasiparticle) bandwidth, and then
the energy scale of the electron quasiparticle band is controlled
by the magnetic interaction $J$, while $\Sigma^{(h)}_{e}$
renormalizes the chemical potential, and therefore plays an
important role in qualitatively determining the positions of peaks
from the doping dependence of the electron spectrum. In this case,
the quasiparticle coherent weight $Z_{F}$ satisfies the following
equation,
\begin{eqnarray}
&Z^{-1}_{F}&=1+{1\over N^{2}}\sum_{{\bf pp'}}\Lambda({\bf k,p,p'})
Z_{F}{B_{{\bf p'}} B_{{\bf p+p'}}\over 4\omega_{{\bf p'}}
\omega_{{\bf p+p'}}}\nonumber \\
&\times& \left ({F_{1}({\bf k,p,p'})\over (\omega_{{\bf p+p'}}-
\omega_{{\bf p'}}-\bar{\xi}_{{\bf p+k}})^{2}}+{F_{2}({\bf k,p,p'})
\over (\omega_{{\bf p'}} -\omega_{{\bf p+p'}}-\bar{\xi}_{{\bf
p+k}})^{2}}\right .\nonumber \\
&+&\left. {F_{3}({\bf k,p,p'})\over (\omega_{{\bf p'}}+
\omega_{{\bf p+p'}}-\bar{\xi}_{{\bf p+k}})^{2}}+{F_{4}({\bf k,p,
p'})\over (\omega_{{\bf p+p'}}+\omega_{{\bf p'}} +\bar{\xi}_{{\bf
p+k}})^{2}}\right ),
\end{eqnarray}
where $F_{1}({\bf k,p,p'})=n_{F}(\bar{\xi}_{{\bf p+k}})[n_{B}
(\omega_{{\bf p'}})-n_{B}(\omega_{{\bf p+p' }})]-n_{B}
(\omega_{{\bf p+p'}})n_{B}(-\omega_{{\bf p'}})$, $F_{2} ({\bf
k,p,p'})=n_{F}(\bar{\xi}_{{\bf p+k}})[n_{B}(\omega_{{\bf p'+p}})
-n_{B}(\omega_{{\bf p'}})]-n_{B}(\omega_{{\bf p'}})n_{B}(-
\omega_{{\bf p'+p}})$, $F_{3}({\bf k,p,p'})= n_{F}
(\bar{\xi}_{{\bf p+k}})[n_{B}(\omega_{{\bf p+p'}})-n_{B}
(-\omega_{{\bf p'}})]+n_{B}(\omega_{{\bf p'}})n_{B}(\omega_{{\bf
p+p'}})$, and $F_{4} ({\bf k,p,p'})=n_{F}(\bar{\xi}_{{\bf p+k})}
[n_{B}(-\omega_{{\bf p'}})-n_{B} (\omega_{{\bf p+p'}})]+
n_{B}(-\omega_{{\bf p'}})n_{B} (-\omega_{{\bf p+p'}})$. This
self-consistent equation must be solved simultaneously with other
self-consistent equations \cite{feng2},
\begin{mathletters}
\begin{eqnarray}
\phi_{1}&=&{1\over 2N}\sum_{{\bf k}}\gamma_{{\bf k}}Z_{F}\left
(1-{\rm th}[{1\over 2}\beta\bar{\xi_{{\bf k}}}]\right ),\\
\phi_{2}&=&{1\over 2N}\sum_{{\bf k}}\gamma_{{\bf k}}'Z_{F}\left
(1-{\rm th}[{1\over 2}\beta \bar{\xi_{{\bf k}}}]\right ),\\
\delta &=& {1\over 2N}\sum_{{\bf k}}Z_{F}\left (1-{\rm th}[{1\over
2}\beta \bar{\xi_{{\bf k}}}]\right ),\\
\chi_{1}&=&{1\over N}\sum_{{\bf k}}\gamma_{{\bf k}} {B_{{\bf
k}}\over 2\omega_{{\bf k}}}{\rm coth}
[{1\over 2}\beta\omega_{{\bf k}}],\\
\chi_{2}&=&{1\over N}\sum_{{\bf k}}\gamma_{{\bf k}}'{B_{{\bf
k}}\over 2\omega_{{\bf k}}}{\rm coth}
[{1\over 2}\beta\omega_{{\bf k}}],\\
C_{1}&=&{1\over N}\sum_{{\bf k}}\gamma^{2}_{{\bf k}} {B_{{\bf
k}}\over 2\omega_{{\bf k}}}{\rm coth}
[{1\over 2}\beta\omega_{{\bf k}}],\\
C_{2}&=&{1\over N}\sum_{{\bf k}}\gamma'^{2}_{{\bf k}} {B_{{\bf
k}}\over 2\omega_{{\bf k}}}{\rm coth}  [{1\over 2}
\beta\omega_{{\bf k}}], \\
C_{3}&=&{1\over N}\sum_{{\bf k}}\gamma_{{\bf k}}\gamma_{{\bf k}}'
{B_{{\bf k}}\over 2\omega_{{\bf k}}}{\rm coth} [{1\over 2}
\beta\omega_{{\bf k}}],\\
{1\over 2} &=&{1\over N}\sum_{{\bf k}}{B_{{\bf k}} \over
2\omega_{{\bf k}}}{\rm coth} [{1\over 2}\beta\omega_{{\bf k}}],\\
\chi^{z}_{1}&=&{1\over N}\sum_{{\bf k}}\gamma_{{\bf k}}{B_{z{\bf
k}}\over 2\omega_{z{\bf k}}}{\rm coth}[{1\over 2}\beta
\omega_{z{\bf k}}],\\
\chi^{z}_{2}&=&{1\over N}\sum_{{\bf k}}\gamma_{{\bf k}}'{B_{z{\bf
k}}\over 2\omega_{z{\bf k}}}{\rm coth}[{1\over 2}\beta
\omega_{z{\bf k}}],\\
C^{z}_{1}&=&{1\over N}\sum_{{\bf k}}\gamma^{2}_{{\bf k}} {B_{z{\bf
k}}\over 2\omega_{z{\bf k}}}{\rm coth} [{1\over 2}\beta
\omega_{z{\bf k}}],\\
C^{z}_{3}&=&{1\over N}\sum_{{\bf k}}\gamma_{{\bf k}}\gamma_{{\bf
k}}'{B_{z{\bf k}}\over 2\omega_{z{\bf k}}}{\rm coth} [{1\over 2}
\beta\omega_{z{\bf k}}],
\end{eqnarray}
\end{mathletters}
then all order parameters, decoupling parameter $\alpha$, and
chemical potential $\mu$ are determined by the self-consistent
calculation. In this sense, our above calculations are exact
without using adjustable parameters, in other words, they are
controllable.

With the help of the dressed charge carrier Green's function
$g({\bf k},\omega)$ in Eq. (9) and MF spin Green's function
$D^{(0)}({\bf p},\omega)$ in Eq. (4b), the electron Green's
function in Eq. (3) can be evaluated explicitly as,
\begin{mathletters}
\begin{eqnarray}
G_{{\rm p-type}}({\bf k},\omega)&=&{1\over N}\sum_{{\bf p}}Z_{F}
{B_{{\bf p+k}}\over 2\omega_{{\bf p+k}}}\left ({L_{1}({\bf k,p})
\over\omega+\bar{\xi}_{{\bf p}}-\omega_{{\bf p+k}}} \right .
\nonumber \\
&+&\left .{L_{2}({\bf k,p})\over \omega+ \bar{\xi}_{{\bf p}}+
\omega_{{\bf p+k}}}\right ), \\
G_{{\rm n-type}}({\bf k},\omega)&=&{1\over N}\sum_{{\bf p}}Z_{F}
{B_{{\bf p+k}}\over 2\omega_{{\bf p+k}}}\left ({L_{2}({\bf k,p})
\over \omega- \bar{\xi}_{{\bf p}}-\omega_{{\bf p+k}}}\right .
\nonumber \\
&+&\left . {L_{1}({\bf k,p})\over\omega-\bar{\xi}_{{\bf p}}+
\omega_{{\bf p+k}}} \right ),
\end{eqnarray}
\end{mathletters}
for hole-doped and electron-doped cases, respectively, where
$L_{1}({\bf k,p})=n_{F}(\bar{\xi}_{{\bf p}})+n_{B}(\omega_{{\bf
p+k}})$ and $L_{2}({\bf k,p})=1-n_{F}(\bar{\xi}_{{\bf p}})+n_{B}
(\omega_{{\bf p+k}})$, then the electron spectral function $A({\bf
k},\omega)=-2{\rm Im}G({\bf k}, \omega)$ is obtained from the
above corresponding electron Green's function as,
\begin{mathletters}
\begin{eqnarray}
A_{{\rm p-type}}({\bf k},\omega)&=&2\pi {1\over N}\sum_{{\bf p}}
Z_{F} {B_{{\bf p+k}}\over 2\omega_{{\bf p+k}}}\nonumber \\
&\times&[L_{1}({\bf k,p})
\delta(\omega+\bar{\xi}_{{\bf p}}-\omega_{{\bf p+k}})\nonumber \\
&+& L_{2}({\bf k,p})\delta(\omega+\bar{\xi}_{{\bf p}}+
\omega_{{\bf p+k}})],\\
A_{{\rm n-type}}({\bf k},\omega)&=&2\pi {1\over N}\sum_{{\bf p}}
Z_{F}{B_{{\bf p+k}}\over 2\omega_{{\bf p+k}}}\nonumber \\
&\times&[L_{2}({\bf k,p})
\delta(\omega-\bar{\xi}_{{\bf p}}-\omega_{{\bf p+k}})\nonumber \\
&+& L_{1}({\bf k,p})\delta(\omega- \bar{\xi}_{{\bf p}}+
\omega_{{\bf p+k}})] .
\end{eqnarray}
\end{mathletters}

We are now ready to discuss the electron spectrum and
quasiparticle dispersion in doped cuprates. Since the absolute
values of $t$ and $t'$ are almost same for both hole-doped and
electron-doped cuprates \cite{hybertson}, therefore the commonly
used parameters in this paper are chosen as $t/J=2.5$ and
$t'/J=0.375$ for the hole doping, and $t/J=-2.5$ and $t'/J=-0.375$
for the electron doping. We have performed the calculation for the
electron spectral function in Eq. (13), and the results at the
$[\pi,0]$ point for (a) the hole doping and (b) electron doping,
and at the $[\pi/2,\pi/2]$ point for (c) the hole doping and (d)
electron doping with temperature $T=0.1J$ in the doping
concentration $\delta=0.09$ (solid line), $\delta=0.12$ (dashed
line), and $\delta=0.15$ (dotted line) are plotted in Fig. 1. For
the hole doping, although both positions of the quasiparticle
peaks at the $[\pi,0]$ and $[\pi/2,\pi/2]$ points are below the
Fermi energy, the position of the quasiparticle peak at the
$[\pi/2,\pi/2]$ point is more close to the Fermi energy, which
indicates that the lowest energy states are located at the
$[\pi/2,\pi/2]$ point. In other words, the low energy spectral
weight with the majority contribution to the low-energy properties
of the hole-doped cuprates comes from the $[\pi/2,\pi/2]$ point.
However, only the position of the quasiparticle peak at the
$[\pi,0]$ point for the electron doping is below the Fermi energy,
and in contrast to the hole-doped case, the position of the
quasiparticle peak at the $[\pi/2,\pi/2]$ point is above the Fermi
energy, which mean that the lowest energy states appear at the
$[\pi,0]$ point, i.e., only states around the $[\pi,0]$ point have
the majority contribution to the low-energy properties of the
electron-doped cuprates. These behaviors reflect the electron-hole
asymmetry in hole-doped and electron-doped cuprates. Moreover, the
electron spectrum is doping dependence. The quasiparticle peaks at
the $[\pi,0]$ and $[\pi/2,\pi/2]$ points for the hole doping and
at the $[\pi,0]$ point for the electron doping become sharper,
while the spectral weight of these peaks increases in intensity
with increasing doping. Furthermore, we have also discussed the
temperature dependence of the electron spectrum, and the results
show that the spectral weight is suppressed with increasing
temperatures. Our these results are qualitatively consistent with
the ARPES experimental data
\cite{shen,kim,dessau,wells,armitage,kim1,koralek}.

\begin{figure}[prb]
\epsfxsize=3.5in\centerline{\epsffile{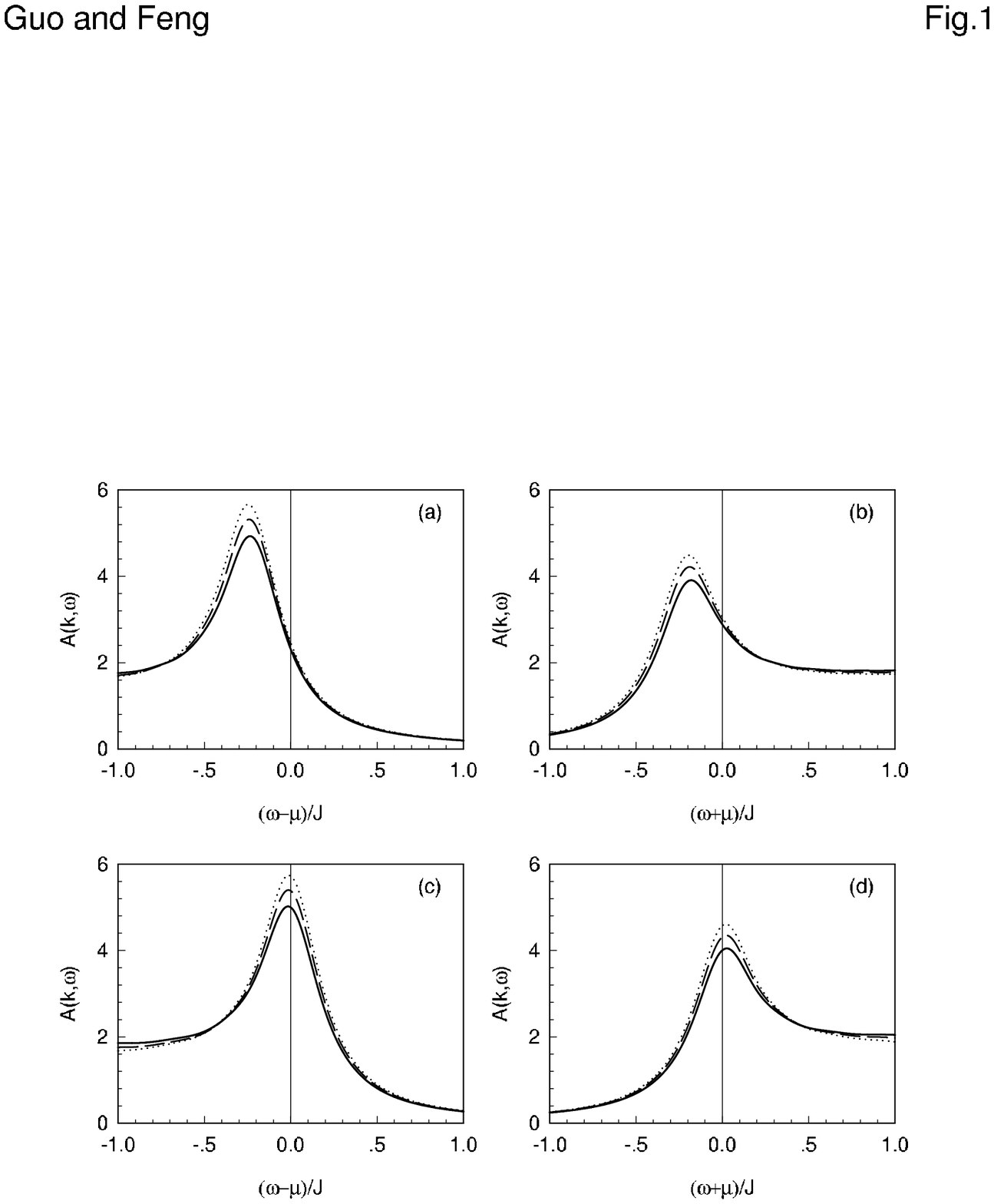}}\caption{The
electron spectral function $A({\bf k},\omega)$ at $[\pi,0]$ point
for (a) the hole doping and (b) electron doping, and at
$[\pi/2,\pi/2]$ point for (c) the hole doping and (d) electron
doping with $T=0.1J$ in $\delta=0.09$ (solid line), $\delta=0.12$
(dashed line), and $\delta=0.15$ (dotted line), where the commonly
used parameters are chosen as $t/J=2.5$ and $t'/J=0.375$ for the
hole doping, and $t/J=-2.5$ and $t'/J=-0.375$ for the electron
doping. }
\end{figure}

\begin{figure}[prb]
\epsfxsize=3.5in\centerline{\epsffile{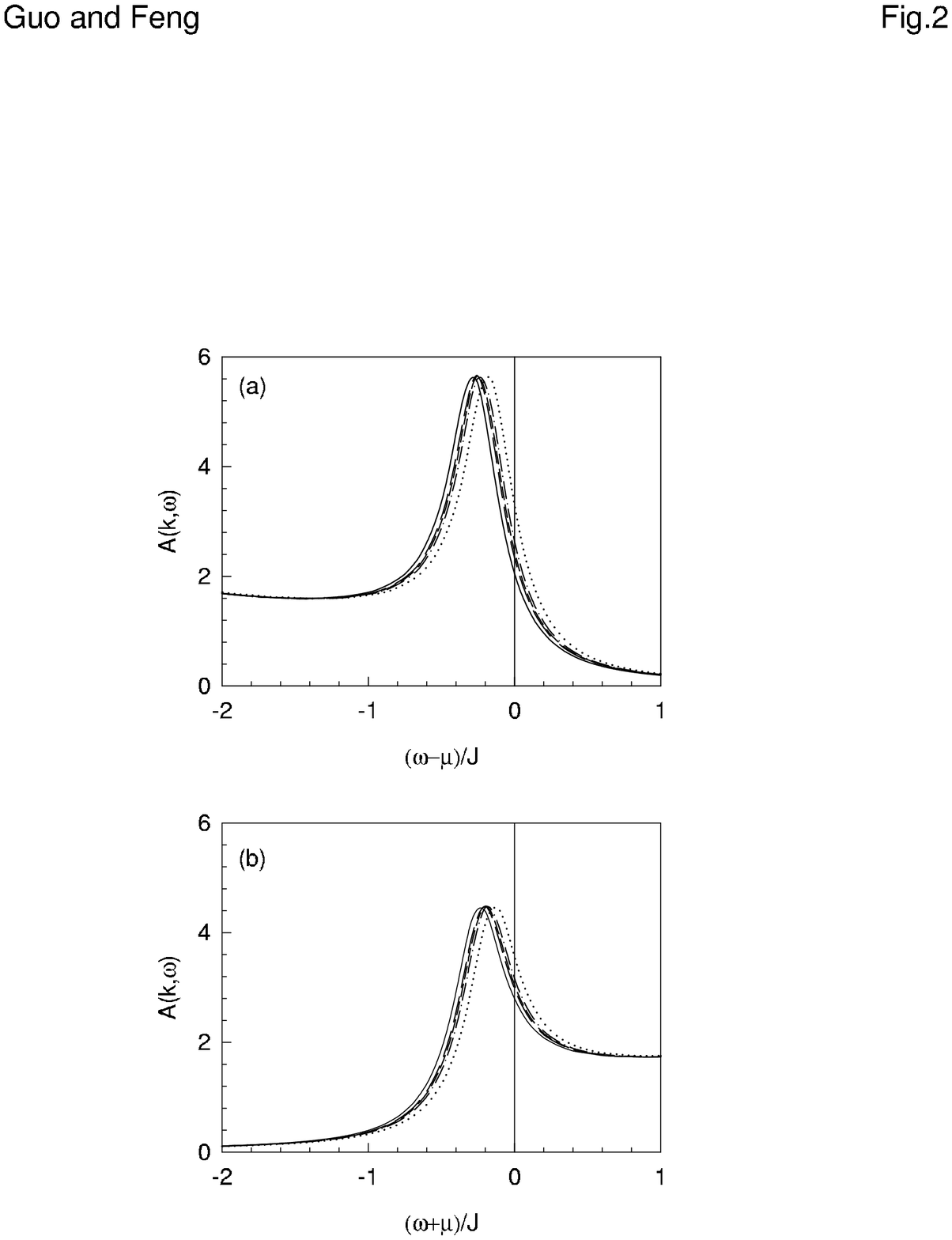}}\caption{The
electron spectral function $A({\bf k},\omega)$ for (a) the hole
doping and (b) electron doping with $T=0.1J$ in $\delta=0.15$ at
$[0.9\pi,0]$ (solid line), $[0.95\pi,0]$ (long dashed line),
$[\pi,0]$ (short dashed line), $[\pi,0.05\pi]$ (dash-dotted line),
and $[\pi,0.1\pi]$ (dotted line) points, where the commonly used
parameters are chosen as $t/J=2.5$ and $t'/J=0.375$ for the hole
doping, and $t/J=-2.5$ and $t'/J=-0.375$ for the electron doping.}
\end{figure}

For a better understanding of the anomalous form of the electron
spectrum $A({\bf k},\omega)$ as a function of energy $\omega$ for
${\bf k}$ in the vicinity of the $[\pi,0]$ point, we have made a
series of calculations for $A({\bf k},\omega)$, and the results
for (a) the hole doping and (b) electron doping with $T=0.1J$ in
$\delta=0.15$ at the $[0.9\pi,0]$ (solid line), $[0.95\pi,0]$
(long dashed line), $[\pi,0]$ (short dashed line), $[\pi,0.05\pi]$
(dash-dotted line), and $[\pi,0.1\pi]$ (dotted line) points are
plotted in Fig. 2. Obviously, the positions of these peaks of the
electron spectral function $A({\bf k},\omega)$ around the
$[\pi,0]$ point are almost not changeable, which leads to the
unusual quasiparticle dispersion around the $[\pi,0]$ point.
Furthermore, the lowest energy peaks are well defined at all
momenta. To show the broad feature in the electron spectrum around
the $[\pi,0]$ point clearly, we plot the positions of the lowest
energy quasiparticle peaks in $A({\bf k},\omega)$ as a function of
momentum along the high symmetry directions for (a) the hole
doping and (b) electron doping with $T=0.1J$ at $\delta=0.15$ in
Fig. 3. For comparison, the corresponding results of the bare
electron dispersion of the $t$-$t'$ model (dotted line), and
experimental results (inset) of the electron dispersion from the
hole-doped cuprate \cite{dessau}
Bi$_{2}$Sr$_{2}$CaCu$_{2}$O$_{8+\delta}$ and electron-doped
cuprate \cite{kim} Nd$_{2-x}$Ce$_{x}$CuO$_{4+\delta}$ are also
shown in Fig. 3. Our these results show that in accordance with
the anomalous property of the electron spectrum in Fig. 2, the
electron quasiparticles around the $[\pi,0]$ point disperse very
weakly with momentum, and then the unusual flat band appears,
while the Fermi energy is only slightly above this flat band, in
qualitative agreement with these obtained from ARPES experimental
measurements on doped cuprates
\cite{shen,kim,dessau,wells,armitage,kim1}.

\begin{figure}[prb]
\epsfxsize=3.5in\centerline{\epsffile{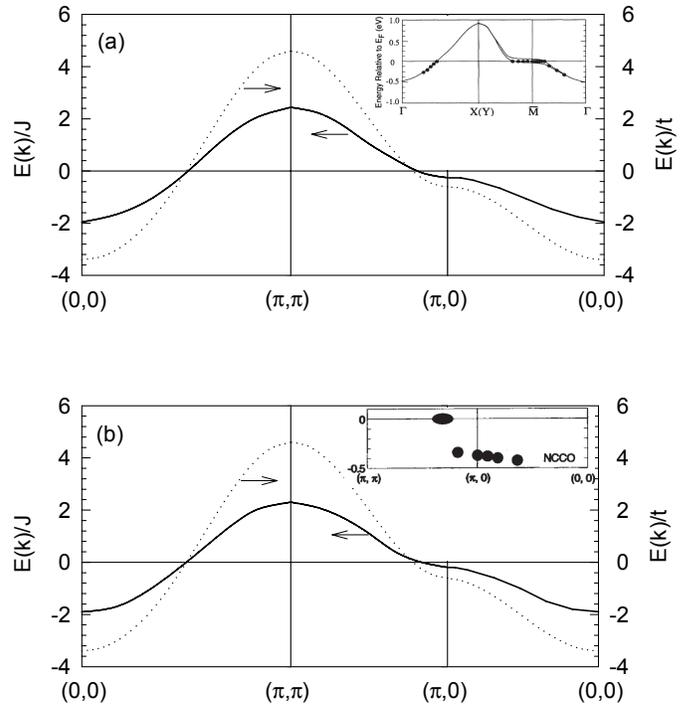}} \caption{The
position of the lowest energy quasiparticle peaks in $A({\bf
k},\omega)$ as a function of momentum for (a) the hole doping with
$t/J=2.5$ and $t'/J=0.375$, and (b) electron doping with
$t/J=-2.5$ and $t'/J=-0.375$, in $T=0.1J$ at $\delta=0.15$. The
dotted line are corresponding results of the bare electron
dispersion of the $t$-$t'$ model. Inset: the corresponding
experimental results of the hole-doped cuprate
Bi$_{2}$Sr$_{2}$CaCu$_{2}$O$_{8+\delta}$ and electron-doped
cuprate Nd$_{2-x}$Ce$_{x}$CuO$_{4+\delta}$ taken from Refs. [12]
and [11], respectively.}
\end{figure}

Since the full electron Green's function (then the electron
spectral function) is obtained beyond the MF approximation by
considering the fluctuation due to the spin pair bubble, therefore
the nature of the electron spectrum is closely related to the
strong coupling between the dressed charge carriers (then electron
quasiparticles) and collective magnetic excitations. This can be
understood from a comparison between the bare electron dispersion
of the $t$-$t'$ model and renormalized electron quasiparticle
dispersion of the $t$-$t'$-$J$ model in Fig. 3. Our results show
that the single-particle hopping in the $t$-$t'$-$J$ model is
strongly renormalized by the magnetic interaction. As a
consequence, the quasiparticle bandwidth is reduced to the order
of (a few) $J$, and therefore the energy scale of the
quasiparticle band is controlled by the magnetic interaction. This
renormalization due to the strong interaction is then responsible
for the unusual electron quasiparticle spectrum and production of
the flat band. On the other hand, although $t'$ does not change
spin configuration, the interplay of $t'$ with $t$ and $J$ causes
a further weakening of the AF spin correlation for the hole
doping, and enhancing the AF spin correlation for the electron
doping \cite{gooding}, which shows that the AF spin correlations
in the electron doping is stronger than these in the hole-doped
side, and leads to the asymmetry of the electron spectrum in
hole-doped and electron-doped cuprates. This is why $t'$ term
plays an important role in explaining the difference between
electron and hole doping. Moreover, our present results also show
that the electron quasiparticle excitations originating from the
dressed holons and spins are due to the charge-spin recombination,
this reflects the composite nature of the electron quasiparticle
excitations, and then the unconventional normal-state properties
in doped cuprates are attributed to the presence of the dressed
charge carriers, spin, and electron quasiparticle excitations.

Finally, we have noted that an obvious weakness of the present
results is that the flat band for the electron doping is not well
below the Fermi energy. In the above calculation, the spin part
has been limited to the MF level, and therefore the spin
fluctuations beyond the MF level is not considered. Since the AF
spin correlations in the electron doping is stronger than these in
the hole doping as mentioned above, it is then possible that the
weakness perhaps due to neglecting the spin fluctuations in the
present case may be cured by considering them, and these and other
related issues are under investigation now.

In summary, we have studied the asymmetry of the electron spectrum
and quasiparticle dispersion in hole-doped and electron-doped
cuprates based on the $t$-$t'$-$J$ model. Our results show that
the quasiparticle dispersions of both hole-doped and
electron-doped cuprates exhibit the flat band around the $[\pi,0]$
point below the Fermi energy. The lowest energy states are located
at the $[\pi/2,\pi/2]$ point for the hole doping, while they
appear at the $[\pi,0]$ point for the electron doping due to the
electron-hole asymmetry. Our results also show that the unusual
behavior of the electron spectrum and quasiparticle dispersion is
intriguingly related to the strong coupling between the electron
quasiparticles and collective magnetic excitations. Within the CSS
fermion-spin theory, we have developed a kinetic energy driven SC
mechanism \cite{feng5}, where the dressed charge carriers interact
occurring directly through the kinetic energy by exchanging the
spin excitations, leading to a net attractive force between the
dressed charge carriers, then the electron Cooper pairs
originating from the dressed charge carrier pairing state are due
to the charge-spin recombination, and their condensation reveals
the SC ground-state. Based on this SC mechanism, we have discussed
the doping and temperature dependence of the electron spectrum of
hole-doped and electron-doped cuprates in the SC-state, and
related theoretical results will be presented elsewhere.

\acknowledgments The authors would like to thank Dr. T.X. Ma and
Dr. Y. Lan for the helpful discussions. This work was supported by
the National Natural Science Foundation of China under Grant Nos.
10125415 and 90403005, and the 973 project from the Ministry of
Science and Technology of China under Grant No. 2006CB601002.

\end{document}